# PM noise of a 40 GHz air-dielectric cavity oscillator

Archita Hati, Craig W. Nelson, B.Riddle and David A. Howe
National Institute of Standards and Technology
Boulder, USA
archita.hati@nist.gov

*Abstract*— **We describe the design of a low-phase-modulated (PM) noise, 40 GHz oscillator that uses a conventional air-dielectric cavity resonator as a frequency discriminator to improve the PM noise of a commercial 10 GHz dielectric resonator oscillator (DRO) frequency multiplied by four. The main features of this design incorporate (1) unloaded cavity quality factor (Q) of 30,000, (2) high coupling coefficient, (3) large carrier suppression by use of interferometric signal processing, (4) large operating signal power of approximately 1 watt (W), and (5) relatively small size. In addition, we report the PM noise of several Ka-band components.**

## INTRODUCTION

Spectrally-pure oscillators are required for millimeter-wave communication and radar systems operating at Ka-band (26.5 to 40 GHz). Designing such oscillators at these high frequencies is quite challenging due to the frequency limitations of active devices. One approach of generating a millimeter-wave reference signal is to simply multiply the frequency of a high-Q quartz oscillator operating at a lower sub-multiple frequency. However, this technique has its own limitations. The best low frequency oscillators are often bulky and costly. Also, frequency multipliers introduce higher phase-modulated (PM) noise at offset frequencies far from the carrier. When a low-noise oscillator is multiplied up to a higher frequency, its noise increases by $10\log(N^2)$, where $N$ is the multiplication factor. The noise of the frequency-multiplied signal is usually higher than the multiplier noise at offset frequencies close to the carrier, but lower at offset frequencies far from the carrier. Therefore, due to the inherent noise of the multiplier, the low PM noise of an oscillator cannot be purely up-converted by simple frequency multiplication.

Some microwave oscillators with the lowest noise employ frequency locking to a high-Q resonance cavity to improve the broadband PM noise [1-5]. The cavity resonator is used primarily as a frequency discriminator. Any improvement of the discriminator phase-shift sensitivity directly translates to lowering the oscillator PM noise. There are several key aspects of controlling the cavity discriminator sensitivity, and the most important of these involves increasing the cavity Q [2, 6]. An effective method of increasing discriminator sensitivity is by suppression of both the carrier signal reflected from the cavity and amplification of the residual noise [2, 3]. The suppression reduces the effective noise temperature of the nonlinear mixer, which acts as a phase detector with enhanced sensitivity. The amount of carrier suppression can be increased by making the effective coupling coefficient into the cavity approach its critical value of unity [2] and also by use of interferometric signal processing [4, 5]. An important aspect of the discriminator sensitivity is that it is proportional to the power of the oscillator signal incident into the cavity [7]. So, by increasing the power of the carrier signal, the discriminator sensitivity can be improved as long as the resonator remains linear, where power level does not change frequency. The purpose of this paper is to study the performance of an air-dielectric cavity resonator that contains all these attributes. These design considerations not only work in the development of a cavity-stabilized oscillator (CSO) of high spectral purity at 40 GHz, but have notable advantages when compared to the usual 10 GHz case. In Section II of this paper, we first provide the phase-noise performance

---

of an assortment of active components at 40 GHz, data that are omitted in most manufacture's datasheets. In Section III, we discuss the design of our compact dielectric-cavity resonator. A description of and phase noise performance for the CSO are respectively given in Section IV and V. Finally, we present a summary in Section VI.

## PM Noise Of Active Components At 40 GHz

The PM noise of devices must be understood before implementing them in a master system. Frequently a component with higher noise is selected that affects the overall performance of the system. The goal of this section is to provide PM noise results for a few selective commercial components at Ka-band, since diminutive or no information is available. We measured the noise of several amplifiers, dividers, and multipliers prior to the design of our 40 GHz oscillator to be discussed in Section III. Each component and its corresponding PM noise are shown in Figs. 1 through 4, respectively. Beginning with Fig. 2, which shows the PM noise of several commercially available amplifiers, Amp1 and Amp3 have roughly similar gain and P1 dB compression, while there is variation of almost 20 dB in the PM noise performance. Similar variation in the noise occurs for the two - dividers shown in Fig. 3. We compared the noise of two dividers: one commercial and the other a custom regenerative divide-by-4 (two divide by-2 in series) [8-9]. In addition, we compared the noise of two frequency multipliers; Fig. 4 shows the noise performance for both. These wide variations in noise performance from one device to another indicate that it is critical to identify the correct components for implementation in a low-noise system for this frequency band.

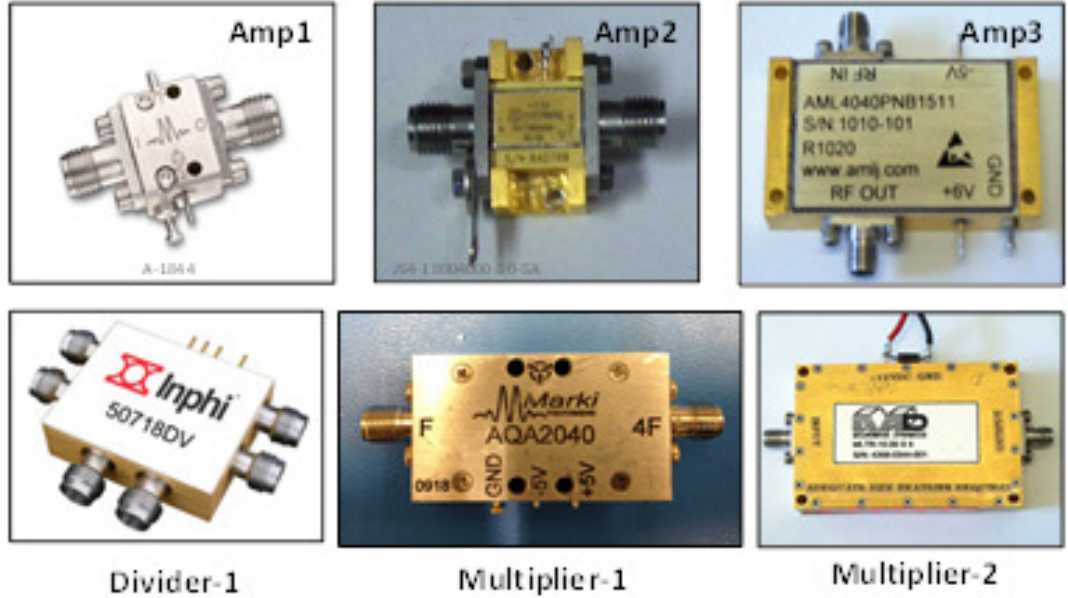

Fig. 1. Images of commercial products used for PM noise measurement at 40 GHz.

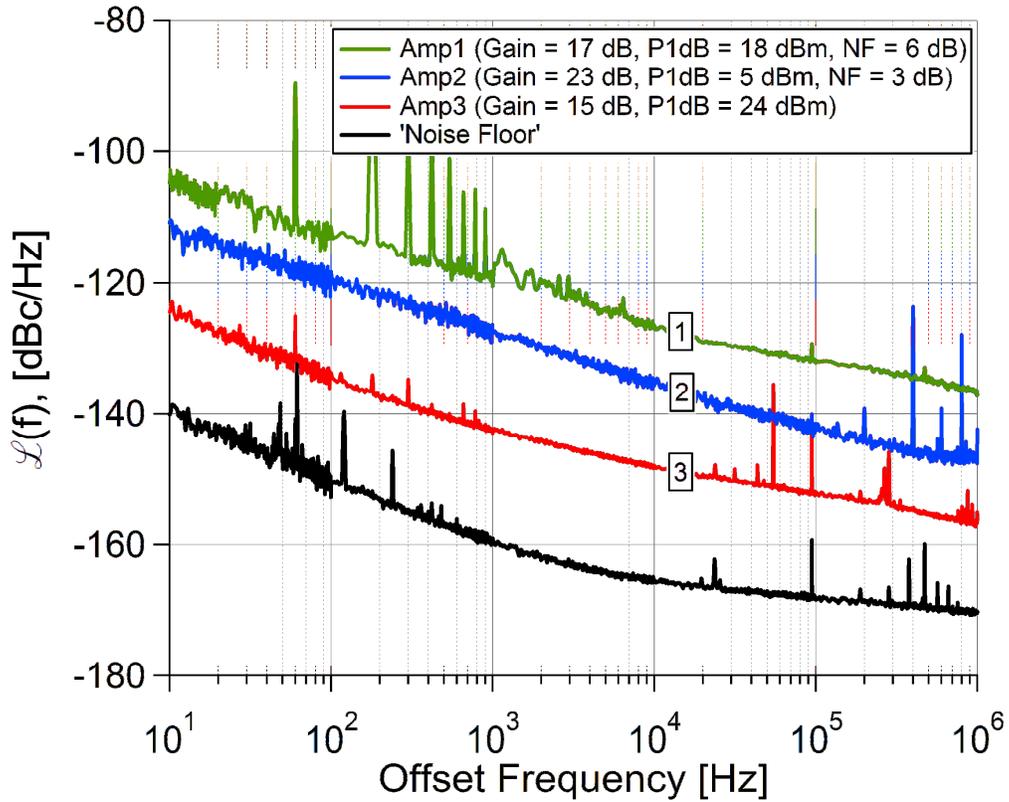

Fig. 2. PM noise for a sample of commercial amplifiers. Carrier frequency = 40 GHz.

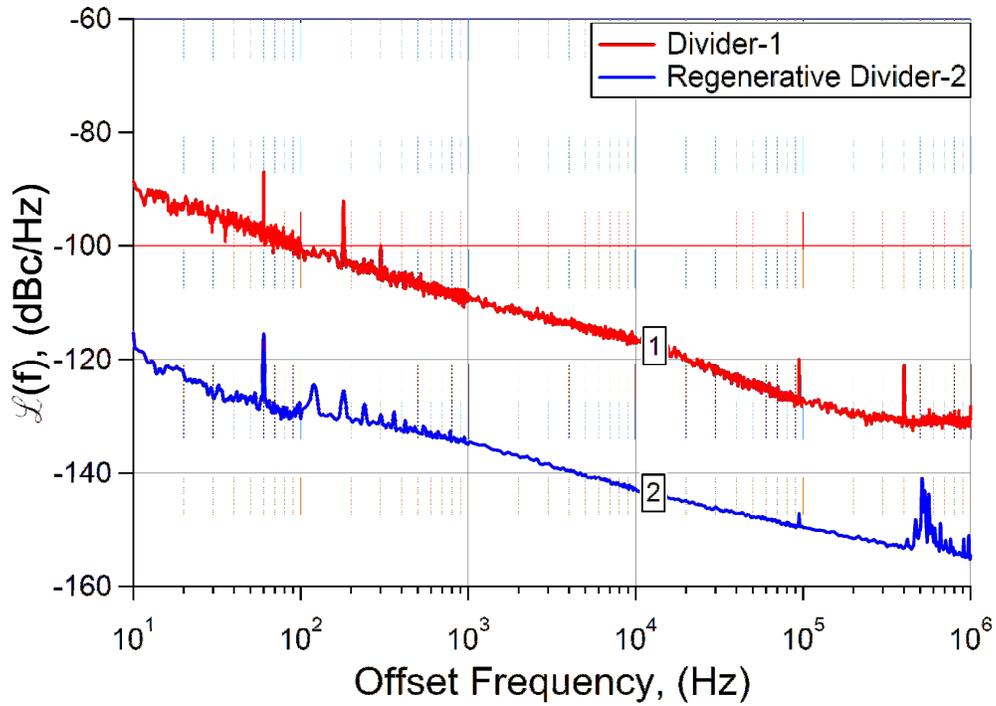

Fig. 3. Input-referred (i.e. 40 GHz) PM noise for a pair of dividers. Divider-1 is a commercial divider shown in Fig. 1. The regenerative divider is custom-built with two divide-by-2 in series. Input frequency = 40 GHz and output frequency = 10 GHz.

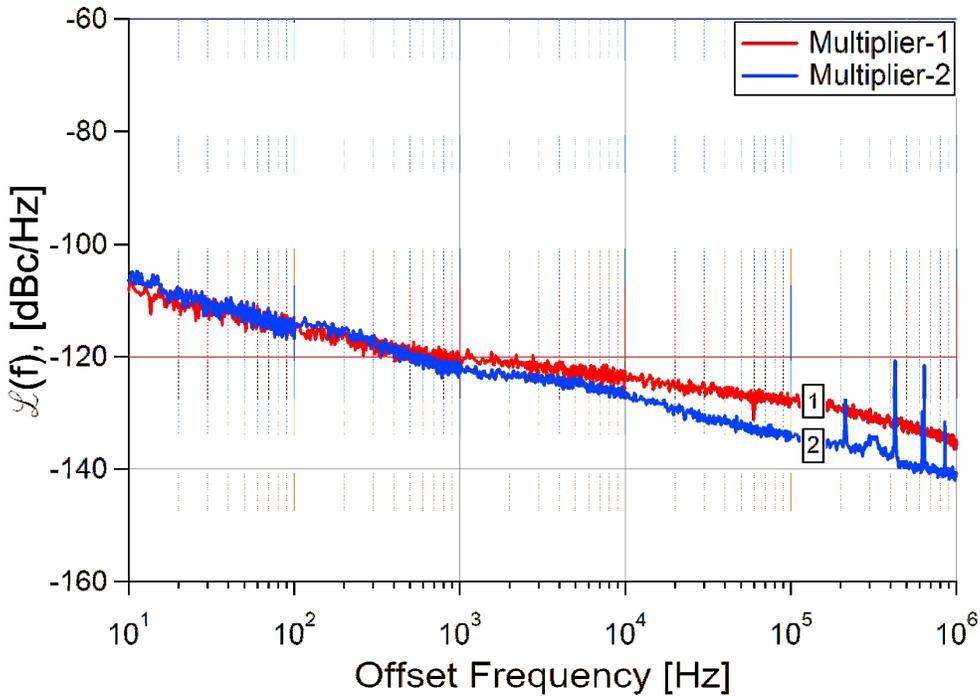

Fig. 4. Output-referred PM noise for a pair of multipliers. Multiplier-1 and Multiplier-2 are shown in Fig. 1. Input frequency =10 GHz, Output frequency = 40 GHz.

COMPACT AIR-DIELECTRIC CAVITY RESONATOR

An important goal of microwave oscillator design is to achieve significant reductions in size, weight, and power (so-called SWaP) without a noise penalty. This paper investigates one strategy of reduced SWaP at state-of-the-art spectral purity. The basic approach used in the past at NIST consists of improving the phase noise of a 10 GHz voltage-controlled oscillator (DRO, yttrium iron garnet (YIG) oscillator, etc.) by use of a high-Q, highly linear air-dielectric 10 GHz cavity as a discriminator [7]. Unloaded Q's of 50,000 to 70,000 are attained for TE023 or TE025 modes, but this moderately high-Q results in a fairly large cavity diameter and height of approximately 8 cm. With a minimal increase in noise, we can substitute a significantly smaller 40 GHz highly-linear air-dielectric cavity as a discriminator. The air-dielectric cylindrical cavity used in the CSO design is operating at TE015 mode, and its inside dimensions are approximately 2 cm ×2 cm . The resonant frequency of this cylindrical cavity is given by [10]

$$f_{res} = \frac{c}{2\pi}\sqrt{\left(\frac{3.832}{a}\right)^2 + \left(\frac{5\pi}{d}\right)^2}, \qquad (1)$$

where, $a$ and $d$ are the radius and length of the cylindrical cavity, respectively. For $2a \approx d \approx 2$ cm, the resonant frequency is approximately 40.08 GHz. The unloaded and loaded quality factor (Q) of the cavity is approximately 30,000 and 17,000, respectively. The cavity is made of aluminum (Al), and the inside surface of the cylinder and end plates are silver-plated and highly polished. The cavity used for the CSO design is shown in Fig. 5a. The signal is coupled to the magnetic field in and out of the cavity by use of coupling probes (loops) on the end plates with their planes aligned with the radial plane of the cylindrical cavity. The typical formulas for the reflection ($S_{11}$) and transmission ($S_{22}$) coefficients at the cavity resonance frequency are given as follows:

$$S_{11} = \frac{1 - \beta_1 + \beta_2}{1 + \beta_1 + \beta_2}, \qquad S_{21} = \frac{2\sqrt{\beta_1\beta_2}}{1 + \beta_1 + \beta_2}. \qquad (2)$$

For input coupling coefficient ($\beta_1$) equal to 0.94 and output coupling coefficient ($\beta_2$) equal to 0.01; the reflected and the transmitted signals out of the cavity are suppressed approximately by 32 dB and 22 dB, respectively, as shown in Fig. 5.

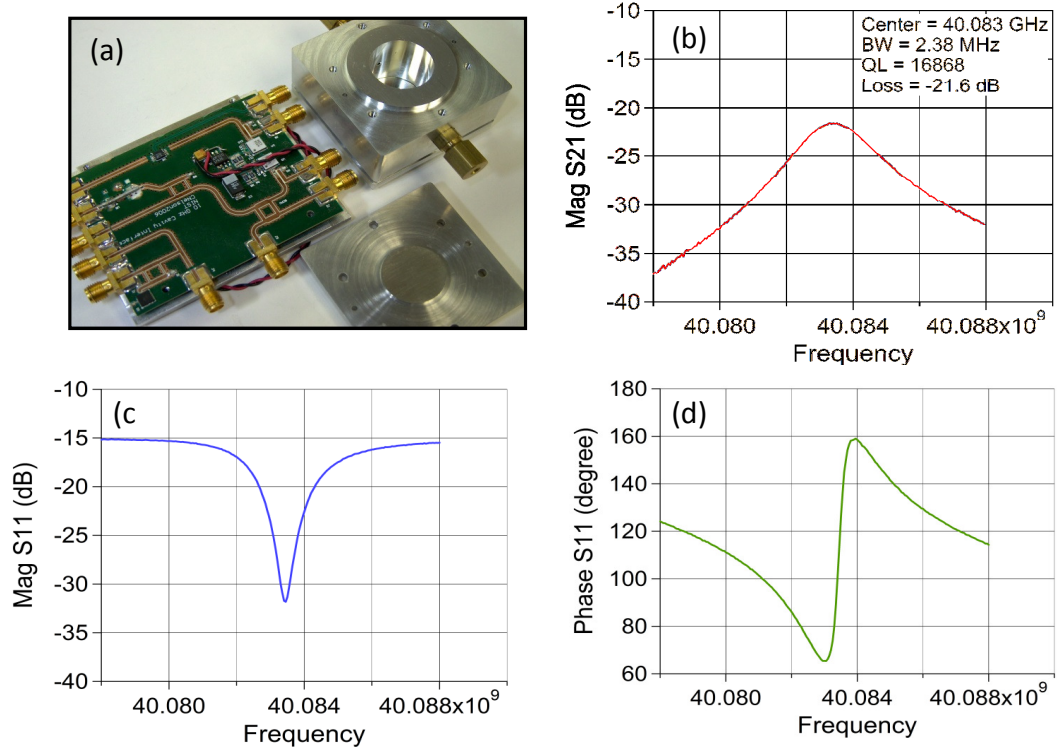

Fig. 5. (a) Air-dielectric cylindrical Al cavity resonator at 40 GHz. The diameter ($2a$) and the length ($d$) of the cavity are approximately 2 cm. (b) Measured data of $|S_{21}|$ of a 40 GHz cavity. (c) Measured data of $|S_{11}|$ of a 40 GHz cavity. (d) Measured phase of $S_{11}$ for a 40 GHz cavity.

### DESCRIPTION OF THE CSO

Fig. 6 shows the block diagram of the CSO at 40 GHz. It consists of a DRO (dielectric resonator oscillator) at 10 GHz whose free-running PM noise is approximately -112 dBc/Hz at a 10 kHz offset frequency. The output of the DRO is first multiplied by a factor of four and amplified to 1 W, by use of a power amplifier. The amplified signal is then applied to the input coupling port of the discriminator cavity through a circulator. The reflected signal out of the cavity exits port 'c' of the circulator and is already highly suppressed because the cavity coupling is nearly critical. A portion of the input signal is added out of phase with the reflected signal to further suppress the carrier (to about -50 dBm). This constitutes the so-called interferometric signal processing. The suppressed-carrier signal is then amplified by use of a low-noise amplifier (gain = 44 dB, noise figure = 2.8 dB) before being applied to one port of a double balanced mixer (DBM) that acts as a phase detector. Due to the high level of carrier suppression, the amplifier's flicker noise contribution is significantly reduced. The other port of the DBM is a directionally-coupled portion of the input signal, adjusted to be in phase quadrature with the reflected signal. By placing the amplifier before the mixer, the effective noise contribution from the mixer is suppressed by the amplifier gain. The output of the DBM is the error voltage that tracks the frequency fluctuations of the DRO relative to the cavity. This error voltage is applied to the voltage-control tuning input of the DRO through the servo amplifier to stabilize its frequency. Figure 7 shows a typical error signal at the output of the DBM versus the frequency difference between the resonance frequency of the cavity and the 10 GHz DRO signal multiplied by 4. This slope, which is at the mid-point of the resonator discriminator curve, is approximately 100 mV/kHz.

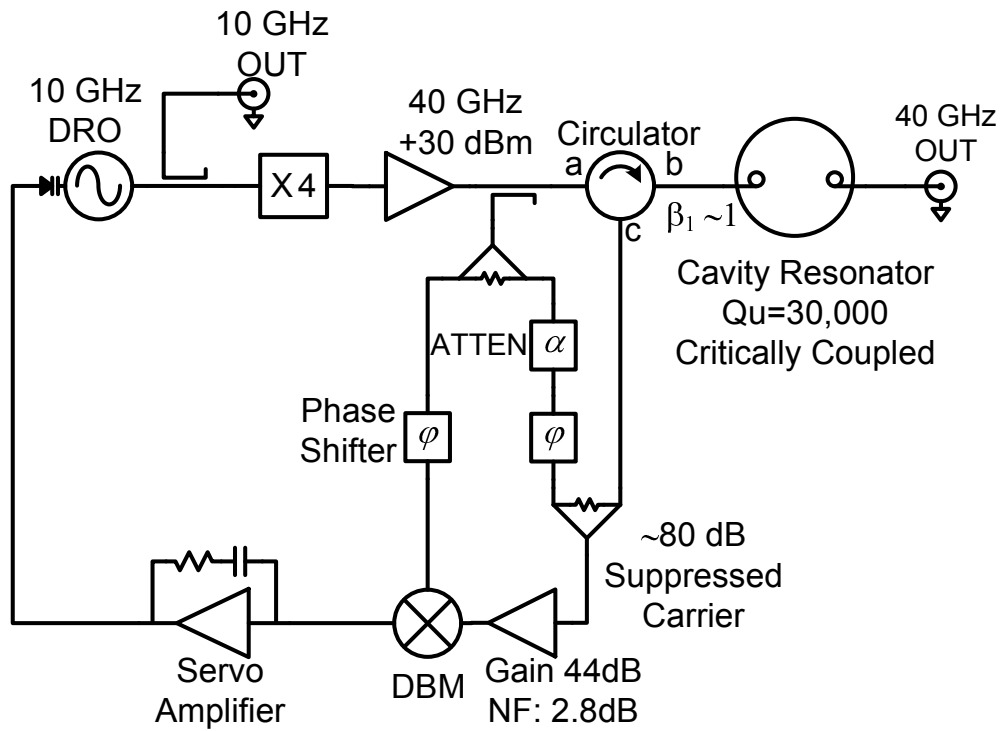

Fig. 6. Block diagram of the cavity stabilized oscillator (CSO) at 40 GHz. The 10 GHz DRO is phase-locked to the phase-transfer function of the air-cavity resonance. DBM is double-balanced mixer, ATTEN is 'attenuator' and $\beta_1$ and $\beta_2$ are respectively the input and output coupling coefficients.

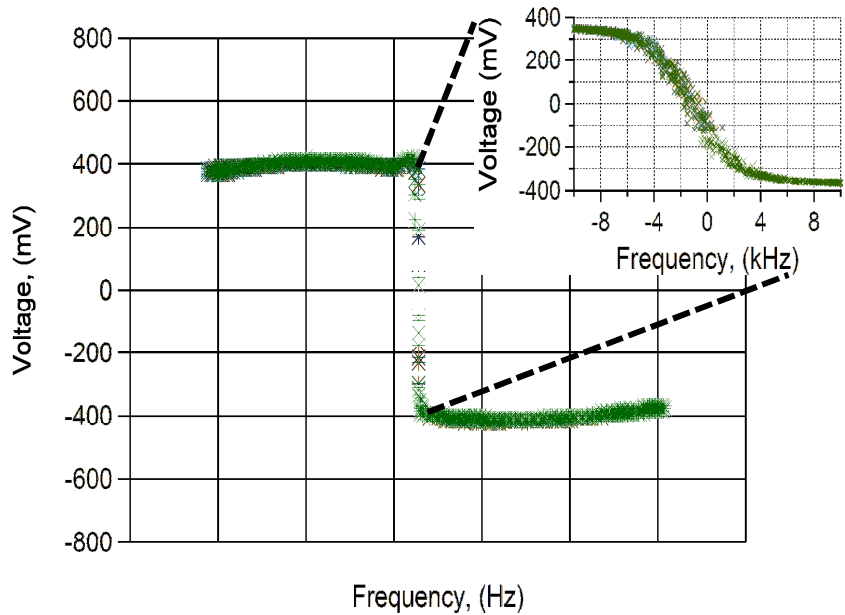

Fig.7. Typical error signal from the double-balanced mixer (DBM) in the cavity discriminator of Fig. 6. The inset corresponds to a slope of 100 mV/kHz approximately for the resonator discriminator curve.

PHASE NOISE RESULTS

Fig. 8 shows the PM noise of a 40 GHz CSO constructed with an aluminum air-dielectric cavity designed for the candidate mode TE015 with an unloaded Q of about 30,000. The PM noise of the free-running DRO is also shown, with the CSO demonstrating almost a 30 dB improvement in the PM noise of the free-running DRO. The noise is measured at 10 GHz at the input of the ×4 multiplier due to the unavailability of a 40 GHz reference oscillator, which has either comparable or better PM noise than the CSO. The final results shown in Fig. 8 are normalized to 40 GHz. There are two drawbacks of measuring the PM noise at the input of the multiplier at 10 GHz. First, the measured noise at 10 GHz is limited by the ×4 multiplier noise. Any corrections of the 40 GHz signal that are lower than the multiplier noise cannot be detected at the 10 GHz output. Second, any improvements to the 40 GHz signal far from the carrier offsets due to the passive filtration of the cavity cannot be observed. There are strategies that can be used to reduce the ×4 multiplier noise, hence reduce the DRO noise, given that the multiplier's output is phase stabilized by the overall CSO scheme. The simplest strategy would be to simply replace the "DRO ×4" with a single low-noise 40 GHz voltage-controlled oscillator (VCO). At lower offset frequencies the PM noise is degraded primarily because of the environmental vibration sensitivity of the 40 GHz resonator.

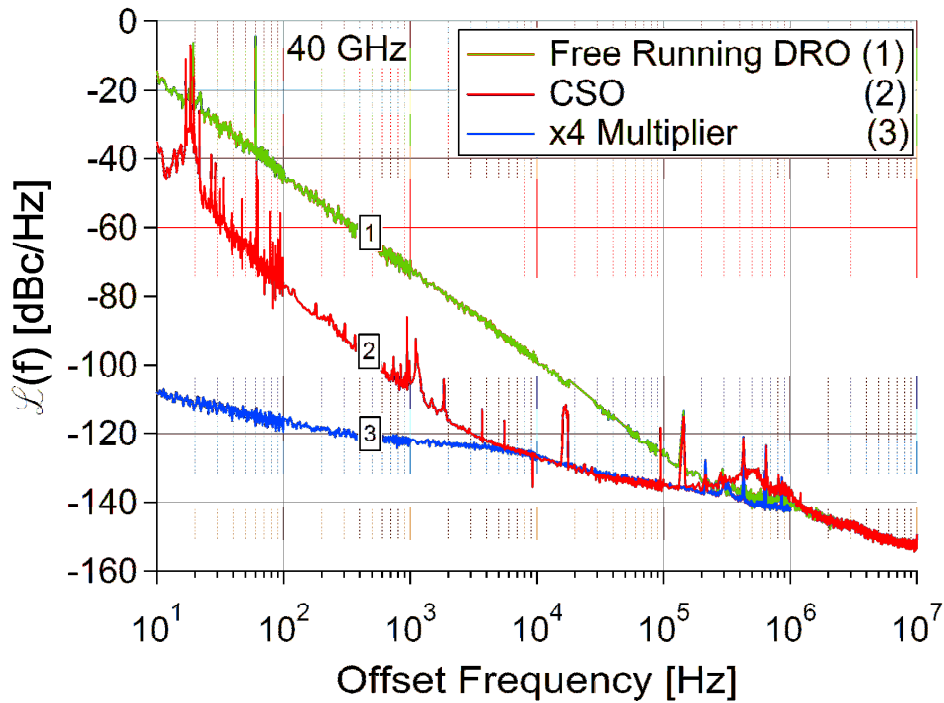

Fig. 8. PM noise of a cavity stabilized oscillator (CSO), the results are normalized to 40 GHz. Notice the significant noise reduction of the free-running DRO noise out to beyond the 100 kHz offset frequency.

CONCLUSIONS

We reported measurements of the PM noise of several commercially available Ka-band components. We observed wide variations in the noise performance among devices, indicating how important it is to correctly identify a component for implementing a low-noise system in this frequency band. We also reported performance of a low-PM noise 40 GHz CSO using an air-dielectric cavity resonator as a frequency discriminator. The cavity in TE015 mode has an unloaded Q of 30,000. The PM noise of the CSO at 10 kHz is -128 dBc/Hz and is entirely limited by the multiplier noise. In the future we plan to

- use a 40 GHz VCO instead of a 10 GHz DRO and a ×4 multiplier
- control the cavity temperature and use vibration isolation to improve close-to-the carrier noise

- use an ultra-stiff ceramic cavity resonator to improve vibration sensitivity of the oscillator.

ACKNOWLEDEMENTS

The authors thank Justin Lanfranchi for the construction and noise measurement of 40 GHz regenerative divide-by-4 circuit, and Stefania Römisch and Jeff Jargon for useful discussion and suggestions. We also thank Danielle Lirette, Mike Lombardi and David Smith for carefully reading and providing comments on this manuscript.